\documentclass[12pt]{amsart}
\usepackage{amssymb,latexsym}
\theoremstyle{plain}
\numberwithin{equation}{section}
\newtheorem{theorem}{Theorem}
\newtheorem{lemma}{Lemma}
\newtheorem{proposition}{Proposition}
\newtheorem{corollary}{Corollary}

\newtheorem{notation}{Notation}

\newtheorem{definition}{Definition}
\begin{document}

AccBou(TRHPWN).tex

\title[ Fock representation of the RHPWN and $w_{\infty}$ algebras]{Fock representation of the renormalized higher powers of white noise and the Virasoro--Zamolodchikov--$w_{\infty}$ $*$--Lie algebra}

\author{Luigi Accardi}
\address{Centro Vito Volterra, Universit\'{a} di Roma Tor Vergata\\
            via Columbia  2, 00133 Roma, Italy}
\email{volterra@volterra.mat.uniroma2.it}
\urladdr{http://volterra.mat.uniroma2.it}
\author{Andreas Boukas}
\address{Department of Mathematics and Natural Sciences, American College of Greece\\Aghia Paraskevi 15342, Athens, Greece}
\email{andreasboukas@acgmail.gr}



\date{\today}

\begin{abstract} The identification of the $*$--Lie algebra of the renormalized higher powers of white noise (RHPWN) and the analytic continuation of the second quantized  Virasoro--Zamolodchikov--$w_{\infty}$ $*$--Lie algebra of conformal field theory and high-energy physics, was recently established  in \cite{id} based on results obtained in \cite{ABIDAQP06} and \cite{ABIJMCS06}. In the present paper we show how the RHPWN Fock kernels must be truncated  in order to be positive definite and we obtain a Fock representation of the two algebras. We show that the truncated renormalized higher powers of white noise (TRHPWN) Fock spaces of order $\geq 2$ host the continuous binomial and beta processes.
\end{abstract}

\maketitle

\section{The RHPWN and Virasoro--Zamolodchikov--$w_{\infty}$ $*$--Lie algebras}

 Let $a_t$ and $a_s^{\dagger}$ be the standard boson white noise functionals with commutator

\[
\lbrack a_t,a_s^{\dagger}\rbrack=\delta(t-s)\cdot 1
\]

\noindent where $\delta$ is the Dirac delta function.  As shown in \cite{ABIDAQP06} and \cite{ABIJMCS06},  using the renormalization

\begin{equation}\label{r1}
\delta^l(t-s)=\delta(s)\,\delta(t-s),\,\,\,\,\,l=2,3,....
\end{equation}

\noindent for the higher powers of the Dirac delta function  and  choosing test functions \linebreak $f:\mathbb{R}\rightarrow \mathbb{C}$ that vanish at zero,  the symbols 

\begin{equation}\label{orig}
B_k^n(f)=\int_{\mathbb{R}}\,f(s)\,{a_s^{\dagger}}^n\,a_s^k\,ds\,\,\,;\,\,\,\,n,k\in\{0,1,2,...\}
\end{equation}

\noindent with involution

\begin{equation}\label{inv}
\left(B_k^n(f)\right)^{*} = B_n^k(\bar f)
\end{equation}

\noindent and

\begin{equation}\label{0}
B_0^0(f)=\int_{\mathbb{R}}\,f(s)\,ds
\end{equation}

\noindent  satisfy the RHPWN commutation relations 

\begin{equation}\label{r2}
\lbrack B^n_k( g),B^N_K(f) \rbrack_{RHPWN}:= \left( k\,N- K\,n     \right)\, B^{n+N-1}_{k+K-1}( g f)
\end{equation}

\noindent where for $n<0$ and/or $k<0$ we define $B_k^n(f):=0$.  Moreover, for $n, N\geq2$ and $k, K\in\mathbb{Z}$ the white noise operators 

\[
\hat{B}_k^n(f):=
\int_{\mathbb{R}}\,f(t)\,e^{ \frac{k}{2}(a_t- a_t^{\dagger})}
\left(\frac{ a_t+ a_t^{\dagger}}{2}\right)^{n-1} \,  
e^{ \frac{k}{2}(a_t- a_t^{\dagger})}\,dt
\]

\noindent satisfy the commutation relations 

\begin{equation}\label{com}
\lbrack \hat B^n_k(g) , \hat B^N_K(f) \rbrack_{w_{\infty}} :=\left(   (N-1)\,k-(n-1)\,K  \right) \,\hat B^{n+N-2}_{k+K}(g\,f)
\end{equation}

\noindent of the Virasoro--Zamolodchikov--$w_{\infty}$ Lie algebra of conformal field theory  with involution

\[
\left( \hat{B}^n_k (f)\right)^*= \hat{B}^n_{-k}(\bar f)
\]

\noindent In particular, for $n=N=2$ we obtain

\[
\lbrack \hat B^2_k(g) , \hat B^2_K (f)\rbrack_{w_{\infty}}  =\left(k-K  \right) \,\hat B^{2}_{k+K}(gf)
\]

\noindent which are the commutation relations of the Virasoro algebra.  The analytic continuation $\{\hat{B}^n_z(f)\,;\,n\geq 2,z\in\mathbb{C}\}$ of the Virasoro--Zamolodchikov--$w_{\infty}$ Lie algebra, and the RHPWN Lie algebra  with commutator $\lbrack \cdot, \cdot \rbrack_{RHPWN}$  have recently been identified (cf. \cite{id}) thus bridging quantum probability with conformal field theory and high-energy physics. 

\begin{notation}\label{not}
In what follows, for all integers $n,k$ we will use the notation $B^n_k:=B^n_k(\chi_I)$  where $I$ is some fixed subset of  $\mathbb{R}$ of finite measure $\mu:=\mu(I)>0$.
\end{notation}

\section{The  action of the RHPWN operators on the Fock  vacuum vector $\Phi$ }

\subsection{Definition of the RHPWN action  on the Fock  vacuum vector $\Phi$ }

 To formulate a reasonable definition of the action of the RHPWN operators on $\Phi$, we go  to the level of white noise.

\begin{lemma}\label{stirling} For all $t\geq s\geq 0$ and $n\in\{0,1,2,...\}$

\[
(a_t^{\dagger})^n\,(a_s)^n=\sum_{k=0}^{n}\,s_{n,k}\,(a_t^{\dagger}\,a_s)^k\,\delta^{n-k}(t-s)
\]

\noindent where $s_{n,k}$ are the Stirling numbers of the 
first  kind with $s_{0,0}=1$ and $s_{0,k}=s_{n,0}=0$ for all $n,k\geq1$.
\end{lemma}

\begin{proof}  As shown in \cite{ABCOSA06}, if $\lbrack b,b^{\dagger}\rbrack= 1$ then 

\begin{equation}\label{a1}
(b^{\dagger})^k\,(b)^k=\sum_{m=0}^k\,s_{k,m}\,(b^{\dagger}\,b)^m
\end{equation}

\noindent For fixed $t,s\in\mathbb{R}$ we define $b^{\dagger}$ and $b$ through 

\begin{equation}\label{a2}
\delta(t-s)^{1/2}\,b^{\dagger}=a^{\dagger}_t,\,\,\mbox{and}\,\,\delta(t-s)^{1/2}\,b=a_s
\end{equation}

\noindent Then $[b,b^{\dagger}]=1$ and the result follows by substituting (\ref{a2}) into (\ref{a1}).
\end{proof}

\begin{proposition}\label{wn} For all integers $n\geq k\geq0$ and for all test functions $f$

\begin{equation}\label{wnformula}
B^n_k(f)=\int_{\mathbb{R}}\,f(t)\,(a_t^{\dagger})^{n-k}\,(a_t^{\dagger}\,a_t)^k\,dt
\end{equation}

\end{proposition}

\begin{proof}  For $n\geq k$ we  can write

\[
(a_t^{\dagger})^n\,(a_s)^k=(a_t^{\dagger})^{n-k}\, (a_t^{\dagger})^k\,(a_s)^k 
\]

\noindent Multiplying both sides by $f(t)\,\delta(t-s)$ and then taking $\int_{\mathbb{R}}\,\int_{\mathbb{R}}...ds\,dt$ of both sides of the resulting equation we obtain

\[
\int_{\mathbb{R}}\,\int_{\mathbb{R}}\,f(t)\,(a_t^{\dagger})^n\,(a_s)^k\,\delta(t-s) \,ds\,dt=\int_{\mathbb{R}}\,\int_{\mathbb{R}}\,f(t)\,(a_t^{\dagger})^{n-k}\, (a_t^{\dagger})^k\,(a_s)^k\,\delta(t-s) \,ds\,dt
\]

\noindent which, after applying (\ref{orig}) to its left  and Lemma \ref{stirling} to its right hand side, yields

\begin{eqnarray*}
B^n_k(f)&=&\sum_{m=0}^{k}\,s_{k,m}\,\int_{\mathbb{R}}\,\int_{\mathbb{R}}\,f(t)\,(a_t^{\dagger})^{n-k}\,(a_t^{\dagger}\,a_s)^m\,\delta^{k-m+1}(t-s)\,ds\,dt\\
&=& s_{k,k}\,\int_{\mathbb{R}}\,\int_{\mathbb{R}}\,f(t)\,(a_t^{\dagger})^{n-k}\,(a_t^{\dagger}\,a_s)^k\,\delta(t-s)\,ds\,dt\\
&+&\sum_{m=0}^{k-1}\,s_{k,m}\,\int_{\mathbb{R}}\,\int_{\mathbb{R}}\,f(t)\,(a_t^{\dagger})^{n-k}\,(a_t^{\dagger}\,a_s)^m\,\delta(s)\,\delta(t-s)\,ds\,dt\\
&=& s_{k,k}\,\,\int_{\mathbb{R}}\,f(t)\,(a_t^{\dagger})^{n-k}\,(a_t^{\dagger}\,a_t)^k\,dt\,+\,0\\
&=&\int_{\mathbb{R}}\,f(t)\,(a_t^{\dagger})^{n-k}\,(a_t^{\dagger}\,a_t)^k\,dt
\end{eqnarray*}

\noindent where we have used the renormalization rule (\ref{r1}), $f(0)=0$, and $ s_{k,k}=1$.  
\end{proof}

\begin{proposition}\label{asig}  Suppose that for all $n,k\in\{0,1,2,...\}$ and  test functions $f$,  

\begin{equation}\label{nat2}
B^n_k(f)\,\Phi:=
\begin{cases}
0 &\text{if $n<k$ or $n\cdot k<0$}\\
B^{n-k}_0(f\,\sigma_k)\,\Phi &\text{if $n>k\geq0$}\\
\int_{\mathbb{R}}\,f(t)\,\rho_k(t)\,dt\,\Phi &\text{if $n= k$}
\end{cases}
\end{equation}

\noindent where  $\sigma_k$  and   $\rho_k$  are  complex valued functions.  Then for all $n\in\{0,1,2,...\}$

\begin{equation}\label{arec5}
\sigma_n={\sigma}_1^n
\end{equation} 

\noindent and

\begin{equation}\label{arec6}
\rho_n=\frac{\sigma_1^n }{n+1}
\end{equation}

\end{proposition}

\begin{proof}   By (\ref{nat2}) and (\ref{orig}) for $k=0$,  and by  (\ref{0}) for $n=k=0$ it follows that $\sigma_0=\rho_0=1$. For $n\geq1$ we have

\begin{eqnarray*}
\langle B^n_0(f)\Phi, B^{n+1}_1(g)\Phi \rangle&=&\langle B^n_0(f)\Phi, B^{n}_0(g\,\sigma_1)\Phi \rangle\\
&=& \langle \Phi,B^0_n(\bar f)\,B^{n}_0(g\,\sigma_1)\,\Phi\rangle\\
&=&\langle \Phi,(B^{n}_0(g\,\sigma_1)\,B^0_n(\bar f)+ \lbrack B^0_n(\bar f),B^{n}_0(g\,\sigma_1)\rbrack)\,\Phi\rangle\\
&=&\langle \Phi,(0+ n^2\, B^{n-1}_{n-1}(\bar f\,g\,\sigma_1)\rbrack)\,\Phi\rangle\\
&=& n^2\, \int_{\mathbb{R}}\,\rho_{n-1}(t)\,\sigma_1(t)\,\bar f(t)\,g(t)\,dt
\end{eqnarray*}

\noindent and also

 \begin{eqnarray*}
\langle B^n_0( f)\Phi, B^{n+1}_1(g)\Phi \rangle&=& \langle \Phi,B^0_n(\bar f)\,B^{n+1}_1(g)\,\Phi\rangle\\
&=&\langle \Phi,(B^{n+1}_1(g)\,B^0_n(\bar f)+ \lbrack B^0_n(\bar f),B^{n+1}_1(g)\rbrack)\,\Phi\rangle\\
&=&\langle \Phi,(0+ n\,(n+1)\,B^{n}_n(\bar f\,g)\rbrack)\,\Phi\rangle\\
&=&n\,(n+1)\,\int_{\mathbb{R}}\,\rho_{n}(t)\,\bar f(t)\,g(t)\,dt
\end{eqnarray*}

\noindent i.e.,  for all test functions $h$

\[
n^2\, \int_{\mathbb{R}}\,\rho_{n-1}(t)\,\sigma_1(t)\,h(t)\,dt=n\,(n+1)\,\int_{\mathbb{R}}\,\rho_{n}(t)\,h(t)\,dt
\]

\noindent  which implies that

\begin{equation}\label{arec}
\rho_n=\frac{n}{n+1}\,\sigma_1\,\rho_{n-1}=...=\frac{\sigma_1^n }{n+1}
\end{equation}

\noindent thus proving (\ref{arec6}).  Similarly,

\begin{eqnarray*}
\int_{\mathbb{R}}\,\rho_{n}(t)\,f(t)\,g(t)\,dt&=&\langle \Phi, B^{n}_{n}(f\,g)\,\Phi \rangle=\frac{ 1}{n+1} \langle \Phi, [B^{n-1}_{n}(f),B^2_1(g)]\,\Phi \rangle\\
&=&\frac{ 1}{n+1} \langle \Phi, (B^{n-1}_{n}(f)\,B^2_1(g)-B^2_1(g)\,B^{n-1}_{n}(f))\,\Phi \rangle\\
&=&\frac{ 1}{n+1} \langle \Phi, B^{n-1}_{n}(f)\,B^2_1(g)\,\Phi \rangle=\frac{ 1}{n+1} \langle B_{n-1}^n(\bar f)\, \Phi, B^2_1(g)\,\Phi \rangle  \\
&=&\frac{ 1}{n+1} \langle B^1_0(\sigma_{n-1}\,\bar f)\, \Phi, B^1_0(\sigma_{1}\,g)\,\Phi \rangle=\frac{ 1}{n+1} \langle \Phi,  B^0_1(\bar{\sigma}_{n-1}\,f)\,\,B^1_0(\sigma_1\,g)\,\Phi \rangle \\
&=&\frac{1}{n+1} \langle \Phi, \lbrack B^0_1(\bar{\sigma}_{n-1}\,f)\,\,B^1_0(\sigma_1\,g)\rbrack\,\Phi \rangle =\frac{1}{n+1} \langle \Phi,  B^0_0(\bar{\sigma}_{n-1}\,f\,\sigma_1\,g)\,\Phi \rangle\\
&=&\frac{ 1}{n+1}\,\int_{\mathbb{R}}\,  \bar{\sigma}_{n-1}(t)\, \sigma_1(t) \,f(t)\,g(t)\,dt
\end{eqnarray*}

\noindent Thus, for all test functions $h$

\[
\int_{\mathbb{R}}\,\rho_{n}(t)\,h(t)\,dt=\frac{ 1}{n+1}\,\int_{\mathbb{R}}\,  \bar{\sigma}_{n-1}(t)\,\sigma_1(t)  \,h(t)\,dt
\]

\noindent therefore

\begin{equation}\label{arec3}
(n+1)\,\rho_{n}= \bar{ \sigma}_{n-1} \,\sigma_1
\end{equation}

\noindent  which combined with (\ref{arec6})  implies

\[
\bar{\sigma}_{n-1}=\sigma_1^{n-1}
\]

\noindent which in turn implies that the $\sigma_n$'s are real and yields (\ref{arec5}).

\end{proof}
 
 In view of the interpretation of $a_t^{\dagger}$ and $a_t$  as creation and annihilation densities respectively, it makes sense to assume that in the definition of the action of $B^n_k$ on $\Phi$ it is only the difference $n-k$ that matters. Therefore we take the function $\sigma_1$ (and thus by (\ref{arec5}) all the $\sigma_n$'s )  appearing in Proposition \ref{asig} to be identically equal to $1$ and we arrive to the following definition of the action of the RHPWN operators on $\Phi$.

\begin{definition}\label{action} For $n,k\in\mathbb{Z}$ and  test functions $f$  

\begin{equation}\label{actionnk}
B^n_k(f)\,\Phi:=
\begin{cases}
0 &\text{if $n<k$ or $n\cdot k<0$}\\
B^{n-k}_0(f)\,\Phi &\text{if $n>k\geq0$}\\
\frac{1}{n+1}\,\int_{\mathbb{R}}\,f(t)\,dt\,\Phi &\text{if $n= k$}
\end{cases}
\end{equation}

\end{definition}

\subsection{The $n$-th order RHPWN  $*$--Lie algebras  $\mathcal{L}_n$}

\begin{definition}\label{L} 

\noindent (i)  $\mathcal{L}_1$ is the $*$--Lie algebra generated by $B^1_0$ and $B^0_1$ i.e., $\mathcal{L}_1$ is the linear span of $\{B^1_0,B^0_1,B^0_0\}$.

\medskip 
 
\noindent (ii) $\mathcal{L}_2$ is the $*$--Lie algebra generated by $B^2_0$ and $B^0_2$ i.e., $\mathcal{L}_2$ is the linear span of $\{B^2_0,B^0_2,B^1_1\}$.

\medskip

\noindent (iii) For $n\in\{3,4,...\}$, $\mathcal{L}_n$ is the $*$--Lie algebra generated by $B^n_0$ and $B^0_n$ through repeated commutations and linear combinations. It  consists of  linear combinations of creation/annihilation operators of the form $B^x_y$ where $x-y=k\,n$ ,  $k\in\mathbb{Z}-\{0\}$, and of number operators $B^{x}_{x}$ with $x\geq n-1$. 
\end{definition}

\subsection{The Fock representation no-go theorem}

 We will show that if the RHPWN action on $\Phi$ is that of  Definition \ref{action}  then the Fock representation no-go theorems of  \cite{ABF06} and \cite{ABIJMCS06}  can be extended to the RHPWN  $*$--Lie algebras  $\mathcal{L}_n$ where $n\geq3$.

\begin{lemma} For all  $n\geq3$ and with the action of the RHPWN  operators on the vacuum vector $\Phi$  given  by Definition \ref{action},  if a Fock space $\mathcal{F}_n$ for $\mathcal{L}_n$ exists then it contains both $B^{n}_0\,\Phi$ and $B^{2n}_0\,\Phi$.
\end{lemma}

\begin{proof} For simplicity we restrict to a single interval $I$ of positive measure $\mu:=\mu(I)$. We have 

\begin{eqnarray*}
B^0_n\,B^n_0\,\Phi&=&(B^n_0\,B^0_n+[B^0_n, B^n_0])\,\Phi=B^n_0\,B^0_n\,\Phi+n^2\,B^{n-1}_{n-1}\,\Phi=0+n^2\,\frac{\mu}{n}\,\Phi=n\,\mu\,\Phi
\end{eqnarray*}

\noindent and

\begin{eqnarray*}
B^0_n\,(B^n_0)^2\,\Phi &=& B^0_n\,B^n_0\,B^n_0\,\Phi=(B^n_0\,B^0_n+n^2\,B^{n-1}_{n-1})\,B^n_0\,\Phi\\
&=&B^n_0\,n\,\mu\,\Phi+n^2\,(B^n_0\,B^{n-1}_{n-1}+[B^{n-1}_{n-1}, B^n_0])\Phi\\
&=&n\,\mu\,B^n_0\,\Phi+n^2\,B^n_0\,\frac{\mu}{n}\,\Phi+n^2\,n\,(n-1)\,B^{2n-2}_{n-2}\,\Phi\\
&=&2\,n\,\mu\,B^n_0\,\Phi+n^3\,(n-1)\,B^n_0\,\Phi\\
&=&(2\,n\,\mu+n^3\,(n-1))\,B^n_0\,\Phi
\end{eqnarray*}

\noindent and also

\begin{eqnarray*}
B^0_n\,(B^n_0)^3\,\Phi&=&(B^n_0\,B^0_n+n^2\,B^{n-1}_{n-1})\,(B^n_0)^2\,\Phi \\
&=&B^n_0\,(2\,n\,\mu+n^3\,(n-1))\,B^n_0\,\Phi+n^2\,(B^n_0\,B^{n-1}_{n-1}+n\,(n-1)\,B^{2n-2}_{n-2})\,B^n_0\,\Phi\\
&=&(2\,n\,\mu+n^3\,(n-1))\,(B^n_0)^2\,\Phi+n^2\,B^n_0\,(B^n_0\,B^{n-1}_{n-1}+n\,(n-1)\,B^{2n-2}_{n-2})\,\Phi\\
&&+n^3\,(n-1)\,(B^n_0\,B^{2n-2}_{n-2}+n\,(n-2)\,B^{3n-3}_{n-3})\,\Phi\\
&=&(2\,n\,\mu+n^3\,(n-1))\,(B^n_0)^2\,\Phi+n^2\,\frac{\mu}{n}\,(B^n_0)^2\,\Phi   +n^3\,(n-1)\,(B^n_0)^2\,\Phi\\
&&+n^3\,(n-1)\,(B^n_0)^2\,\Phi+n^4\,(n-1)\,(n-2)\,B^{2n}_0\,\Phi\\
&=&3\,n\,(\mu+n^2\,(n-1))\,(B^n_0)^2\,\Phi+n^4\,(n-1)\,(n-2)\,B^{2n}_0\,\Phi
\end{eqnarray*}

\noindent Since  $B^0_n\,(B^n_0)^3\,\Phi \in \mathcal{F}_n$ and  $(B^n_0)^2\,\Phi \in \mathcal{F}_n$ it follows that $B^{2n}_0\,\Phi \in \mathcal{F}_n$.

\end{proof}

\begin{theorem}\label{NoGo}  Let  $n\geq3$. If the action of the RHPWN  operators on the vacuum vector $\Phi$ is  given  by  Definition \ref{action}, then $\mathcal{L}_n$ does not admit a Fock representation.
\end{theorem}

\begin{proof} If a  Fock representation of  $\mathcal{L}_n$ existed then we should be able to define inner products of the form

\[
\langle (a\,B^{2n}_0+b\,(B_0^n)^2)\Phi, (a\,B^{2n}_0+b\,(B_0^n)^2)\Phi\rangle
\]

 \noindent  where  $a,b\in \mathbb{R}$ and  the RHPWN operators are defined on the same interval $I$ of arbitrarily small positive measure $\mu(I)$. Using the notation $\langle x\rangle=<\Phi,x\,\Phi\rangle$  this amounts to the positive semi-definiteness of the matrix

\[
A=\left[
\begin{array}{cc}
 \langle B^0_{2n}\,B^{2n}_0\rangle &<B^0_{2n}\,(B^n_0)^2\rangle\\
\langle B^0_{2n}\,(B^n_0)^2\rangle& \langle (B^0_n)^2\,(B^n_0)^2\rangle
\end{array}
\right]
\]

 \noindent Using  (\ref{com}) and Definition \ref{action} we find that 

\begin{equation*}
\langle B^0_{2n}\,B^{2n}_0\rangle = 4\,n^2\,<B^{2n-1}_{2n-1}\rangle= 4\,n^2\,\frac{1}{2\,n}\,\mu(I)=2\,n\, \mu (I)  
\end{equation*}

 \noindent and

\begin{eqnarray*}
\langle B^0_{2n}\,(B^n_0)^2\rangle& =& \langle B_0^{2n}\,\Phi,(B^n_0)^2\,\Phi \rangle=\langle B_n^0\,B_0^{2n}\,\Phi, B^n_0\,\Phi\rangle\\
&=&2\,n^2\,\langle B^{2n-1}_{n-1}\,\Phi, B^n_0\,\Phi\rangle=2\,n^2\,\langle B^{n}_0\,\Phi, B^n_0\,\Phi\rangle\\
&=&2\,n^2\,\langle B_{n}^0\, B^n_0\rangle=2\,n^2\,n^2\,\langle B_{n-1}^{n-1}\rangle\\
&=&2\,n^4\,\frac{1}{n}\,\mu(I)=2\,n^3\,\mu(I)
\end{eqnarray*}

 \noindent and also

\begin{eqnarray*}
\langle (B^0_n)^2\,(B^n_0)^2\rangle& =& \langle B_0^{n}\,\Phi,B_n^0\,(B^n_0)^2\,\Phi \rangle=\langle B^n_0\,\Phi, (B_n^0\,B^n_0)\,B_0^{n}\,\Phi\rangle\\
&=& \langle B_0^{n}\,\Phi, (B^n_0\,B^0_n+n^2\,B^{n-1}_{n-1})\,B_0^{n}\,\Phi \rangle  \\
&=&\langle B_0^{n}\,\Phi, B^n_0\,B^0_n\,B^n_0\,\Phi \rangle+n^2\,\langle B_0^{n}\,\Phi, B^{n-1}_{n-1}\,B^n_0\,\Phi \rangle  \\
&=&\langle B^0_{n}\,B^n_0 \Phi,  B^0_{n}\,B^n_0 \,\Phi \rangle+n^2\,\langle B_0^{n}\,\Phi, (B^n_0\,B^{n-1}_{n-1}+n\,(n-1)\,\,B^{2n-2}_{n-2})\,\Phi \rangle  \\
&=&n^4\,\langle B^{n-1}_{n-1}\Phi, B^{n-1}_{n-1} \Phi \rangle + n\,\mu(I)\,\langle B^{n}_0\,\Phi, B^n_0\,\Phi\rangle+n^3\,(n-1)\,\langle B^n_0\,\Phi, B^{2n-2}_{n-2} \,\Phi \rangle  \\
&=&n^2\,\mu(I)^2+n\,\mu(I)\,\langle B^0_n\,B^n_0 \rangle +n^3\,(n-1)\,\langle B_n^0 B^{2n-2}_{n-2} \rangle  \\
&=&n^2\,\mu(I)^2+n^3\,\mu(I)\,\langle B^{n-1}_{n-1} \rangle +n^4\,(n-1)\,(2\,n-2)\,\langle  B^{2n-3}_{2n-3} \rangle  \\
&=&n^2\,\mu(I)^2+n^2\,\mu(I)^2 +n^4\,(n-1)\,\mu(I)\\
&=&  2\,n^2\,\mu(I)^2 +n^4\,(n-1)\,\mu(I)
\end{eqnarray*}

 \noindent Thus

\[
A=\left[
\begin{array}{cc}
 2\,n\, \mu (I)   &    2\,n^3\,\mu (I)\\
&\\
     2\,n^3\,\mu (I)    & 2\,n^2 \,\mu (I)^2 +n^4\,(n-1)\, \mu (I)
\end{array}
\right].
\]

 \noindent $A$ is a symmetric matrix, so it is positive semi-definite if and only if its minors are non-negative. The minor determinants of $A$ are

\[
d_1= 2\,n\,  \mu (I)
\]

 \noindent  which is always nonnegative, and

\[
d_2= 2\,n^3\,\mu (I)^2  \, (2 \,\mu (I)-n^2-n^3) 
\]

\noindent  which is  nonnegative if and only if

\[
\mu (I) \geq \frac{n^2(n+1)}{2}
\]

\noindent  Thus the interval $I$ cannot be arbitrarily small.
\end{proof}

\section{The $n$-th order truncated RHPWN  (or TRHPWN)  Fock space $\mathcal{F}_n$ }

\subsection{Truncation of the  RHPWN Fock kernels}

  The  generic  element of  the $*$-Lie algebras $\mathcal{L}_n$  of Definition \ref{L}  is $B^n_0$.  All other elements of $\mathcal{L}_n$  are obtained by taking adjoints, commutators, and linear combinations. It thus makes sense to consider ${(B^n_0(f))}^k\,\Phi$ as basis vectors for the $n$-th particle space of the  Fock space  $\mathcal{F}_n$ associated with $\mathcal{L}_n$ . A calculation of the ``Fock kernel''  $\langle (B^n_0)^k\,\Phi, (B^n_0)^k\,\Phi \rangle $  reveals that it is the terms containing $B^{2\,n}_0\,\Phi$ that prevent the kernel from being positive definite. The $B^{2\,n}_0\,\Phi$ terms appear either directly or by applying  Definition \ref{action} to terms of the form  $B^{x}_y\,\Phi$ where $x-y=2\,n$.  Since $\mathcal{L}_1$  and $\mathcal{L}_2$ do not contain $B^2_0$ and $B^4_0$ respectively, that problem exists  for $n\geq 3$ only and the Fock spaces  $\mathcal{F}_1$  and $\mathcal{F}_2$  are actually not truncated.  In what follows we will compute the Fock kernels by applying Definition \ref{action} and  by truncating  ``singular''  terms of the form

\begin{equation}\label{sing}
\langle (B^n_0)^k\,\Phi, (B^n_0)^m\,B^x_y\,\Phi\,\rangle 
\end{equation}

\noindent where $n \, k=n \, m+ x-y$ and $x-y=2\,n$ i.e., $k-m=2$.  This amounts to truncating the action of the  principal  $\mathcal{L}_n$ number operator $B^{n-1}_{n-1}$ on the ``number vectors'' $(B^n_0)^k\,\Phi$,  which by commutation relations (\ref{r2}) and Definition \ref{action} is of the form

\[
B^{n-1}_{n-1}\,(B^n_0)^k\,\Phi=\left(\frac{\mu}{n}+k\,n\,(n-1)\right)\,(B^n_0)^k\,\Phi+\sum_{i\geq1}\,\prod_{ j\geq1}\,c_{i,j}\,B^{\lambda_{i,j}\,n}_0\,\Phi
\]

\noindent ( where for each $i$  not all positive integers $\lambda_{i,j}$  are equal to $1$)  by omitting the \linebreak  $ \sum_{i\geq1}\,\prod_{ j\geq1}\,c_{i,j}\,B^{\lambda_{i,j}\,n}_0\,\Phi $ part.  We thus arrive to the following:

\begin{definition}\label{deftnv} For integers $n\geq1$ and $k\geq0$,

\begin{equation}\label{tnv}
B^{n-1}_{n-1}\,(B^n_0)^k\,\Phi:=\left(\frac{\mu}{n}+k\,n\,(n-1)\right)\,(B^n_0)^k\,\Phi
\end{equation}

\noindent i.e., the number vectors $(B^n_0)^k\,\Phi$ are eigenvectors of the principal   $\mathcal{L}_n$ number operator $B^{n-1}_{n-1}$ with eigenvalues $\left(\frac{\mu}{n}+k\,n\,(n-1)\right)$.

\end{definition}

\noindent In agreement with  Definition \ref{action}, for $k=0$ Definition \ref{deftnv} yields $B^{n-1}_{n-1}\,\Phi:= \frac{\mu}{n}\,\Phi$. 

\subsection{Outline of the Fock space construction method}

 We will construct the TRHPWN Fock spaces by using the following method (cf. Chapter 3 of \cite{fein}): 

\medskip

\noindent (i) Compute 

\[
\|(B^n_0)^k\,\Phi \|^2=\langle (B^n_0)^k\,\Phi, (B^n_0)^k\,\Phi \rangle :=\pi_{n,k}(\mu)
\]

\noindent  where $k=0,1,2,...$, $\Phi$ is the RHPWN vacuum vector, and $\pi_{n,k}(\mu)$ is a polynomial in $\mu$ of degree $k$. 

\medskip
\noindent (ii) Using the fact that  if $k\neq m$ then $\langle (B^n_0)^k\,\Phi, (B^n_0)^m\,\Phi \rangle=0$, for $a,b\in\mathbb{C}$ compute

\begin{eqnarray*}
\langle e^{a\, B^n_0}\,\Phi ,e^{b\, B^n_0}\,\Phi \rangle  &=&\sum_{k=0}^{\infty}\,\frac{(\bar a\,b)^k}{(k!)^2}\,\langle (B^n_0)^k\,\Phi, (B^n_0)^k\,\Phi \rangle\\
&=&\sum_{k=0}^{\infty}\,\frac{(\bar a\,b)^k}{k!}\,\frac{\pi_{n,k}(\mu)}{k!}\\
&=&\sum_{k=0}^{\infty}\,\frac{(\bar a\,b)^k}{k!}\,h_{n,k}(\mu)
\end{eqnarray*}

\noindent where 

\begin{equation}\label{h}
h_{n,k}(\mu):=\frac{\pi_{n,k}(\mu)}{k!}
\end{equation}

\medskip
 \noindent (iii) Look for a  function $G_n(u,\mu)$ such that

\begin{equation}\label{G0}
G_n(u,\mu)=\sum_{k=0}^{\infty}\,\frac{u^k}{k!}\,h_{n,k}(\mu)
\end{equation}

\noindent Using the Taylor expansion of $G_n(u,\mu)$ in powers of $u$

\begin{equation}\label{T}
G_n(u,\mu)=\sum_{k=0}^{\infty}\,\frac{u^k}{k!}\frac{\partial^k }{\partial u^k }\,G_n(u,\mu)|_{u=0}
\end{equation}

\noindent by comparing (\ref{T}) and (\ref{G0}) we see that

\begin{equation}\label{G00}
\frac{\partial^k }{\partial u^k }\,G_n(u,\mu)|_{u=0}=h_{n,k}(\mu)
\end{equation}

\noindent Equation (\ref{G00}) plays a fundamental role in the search for $G_n$  in what follows.

\medskip
\noindent  (iv) Reduce to single intervals and  extend to step functions:  For $u=\bar{a}\,b$,  assuming that 

\begin{equation}\label{T)}
G_n(u,\mu)=e^{\mu\,\hat{G}_n(u)}
\end{equation}

\noindent \noindent which is typical for ``Bernoulli moment systems'' (cf. Chapter 5 of \cite{fein} ), equation  (\ref{G0})  becomes

\begin{equation}\label{G}
e^{\mu\,\hat{G}_n(\bar{a}\, b)}=\sum_{k=0}^{\infty}\,\frac{(\bar{a}\, b)^k}{k!}\,h_{n,k}(\mu)
\end{equation}

 \noindent  Take the product of (\ref{G}) over all sets $I$, for test functions $f:=\sum_i\,a_i\,\chi_{I_i}$ and $g:=\sum_i\,b_i\,\chi_{I_i}$  with  $ I_i \cap I_j= \oslash $ for $i\neq j$,  and end up with an expression like

\begin{equation}\label{inner}
e^{\int_{\mathbb{R}}\, \hat{G}_n(f(t)\, g(t))\,dt}=\prod \, \langle e^{a\, B^n_0}\,\Phi ,e^{b\, B^n_0}\,\Phi \rangle 
\end{equation}

\noindent which we take as the definition of the   inner product $\langle \psi_n(f),\psi_n(g)\rangle_n$ of the ``exponential vectors''

\begin{equation}\label{expvect}
\psi_n(f):=\prod_{i}\,e^{a_i\,B^n_0(\chi_{I_i})}\,\Phi
\end{equation}

\noindent  of the TRHPWN Fock space $\mathcal{F}_n$. Notice that $\Phi=\psi_n(0)$. 

\subsection{Construction of the TRHPWN Fock spaces $\mathcal{F}_n$ }

\begin{lemma}\label{ltnv} Let $n\geq 1$ be fixed. Then for all integers $k\geq0$ 

\begin{equation}\label{tnv2}
B^{0}_{n}\,(B^n_0)^{k+1}\,\Phi:=n\,(k+1)\,\left(\mu+k\,\frac{n^2\,(n-1)}{2}\right)\,(B^n_0)^k\,\Phi
\end{equation}

\end{lemma}

\begin{proof} For $k=0$ we have 

\begin{eqnarray*}
B^0_n\,B^n_0\,\Phi&=&(B^n_0\,B^0_n+[B^0_n, B^n_0])\,\Phi=0+n^2\,B^{n-1}_{n-1}\,\Phi\\
&=&n^2\,\frac{\mu}{n}\,\Phi=n\,\mu\,\Phi=n\,(0+1)\,\left(\mu+0\,\frac{n^2\,(n-1)}{2}\right)\,(B^n_0)^0\,\Phi
\end{eqnarray*}

\noindent Assuming (\ref{tnv2}) to be true for $k$ we have

\begin{eqnarray*}
&&\quad B^0_n\,(B^n_0)^{k+2}\,\Phi=(B^0_n\,B^n_0)\,(B^n_0)^{k+1}\,\Phi=(B^n_0\,B^0_n+n^2\,B^{n-1}_{n-1})\,(B^n_0)^{k+1}\,\Phi \\
\noalign{\vskip .12 true in}
&&\quad=B^n_0\,B^0_n\,(B^n_0)^{k+1}\,\Phi+n^2\,B^{n-1}_{n-1}\,(B^n_0)^{k+1}\,\Phi \\
&&\quad=B^n_0\,n\,(k+1)\,\left(\mu+k\,\frac{n^2\,(n-1)}{2}\right)\,(B^n_0)^{k}\,\Phi+n^2\,B^{n-1}_{n-1}\,(B^n_0)^{k+1}\,\Phi\\
&&\quad=\left(n\,(k+1)\,\left(\mu+k\,\frac{n^2\,(n-1)}{2}\right)+n^2\,\left( \frac{\mu}{n}+(k+1)\,n\,(n-1) \right) \right)\,(B^n_0)^{k+1}\,\Phi\\
&&\quad= n\,(k+2)\,\left(\mu+(k+1)\,\frac{n^2\,(n-1)}{2}\right)\,(B^n_0)^{k+1}\,\Phi
\quad
\end{eqnarray*}

\noindent which proves (\ref{tnv2}) to be true for $k+1$ also, thus completing the induction.

\end{proof}

\begin{proposition}\label{kernels} For all $n\geq 1$

\begin{equation}\label{threeormore}
\pi_{n,k}(\mu):=\langle (B^n_0)^k\,\Phi, (B^n_0)^k\,\Phi \rangle=k!\,n^k\,\prod_{i=0}^{k-1}\left(\mu+\frac{n^2\,(n-1)}{2}\,i\right)
\end{equation}

\end{proposition}

\begin{proof}  Let $n\geq1$ be fixed. Define

\[
a_k:=k!\,n^k\,\prod_{i=0}^{k-1}\left(\mu+\frac{n^2\,(n-1)}{2}\,i\right)
\]

\noindent Then 

\[
a_1=n\,\mu
\]

\noindent and for $k\geq1$

\[
a_{k+1}=n\,(k+1)\,\left(\mu+k\,\frac{n^2\,(n-1)}{2}\right)\,a_k
\]

\noindent Similarly, define

\[
b_k:=\langle (B^n_0)^k\,\Phi, (B^n_0)^k\,\Phi \rangle
\]

\noindent Then

\[
b_1=\langle B^n_0\,\Phi, B^n_0\,\Phi \rangle=\langle \Phi,  B^0_n\,B^n_0\,\Phi \rangle=n^2 \langle \Phi, B^{n-1}_{n-1}\,\Phi \rangle=n^2\,\frac{\mu}{n}=n\,\mu
\]

\noindent and for $k\geq1$, using Lemma \ref{ltnv}

\begin{eqnarray*}
b_{k+1}&=&\langle (B^n_0)^k\,\Phi, B^0_n\,(B^n_0)^{k+1}\,\Phi \rangle=n\,(k+1)\,\left(\mu+k\,\frac{n^2\,(n-1)}{2}\right)\,\langle (B^n_0)^k\,\Phi, (B^n_0)^k\,\Phi \rangle\\
&=&n\,(k+1)\,\left(\mu+k\,\frac{n^2\,(n-1)}{2}\right)\,b_k
\end{eqnarray*}

\noindent Thus $a_k=b_k$ for all  $k\geq 1$.

\end{proof}

\begin{corollary}\label{hs} The functions $h_{n,k}$ appearing in
 (\ref{h}) are given by

\begin{equation}\label{hone}
h_{1,k}=\mu^k
\end{equation}

\noindent  and for $n\geq2$

\begin{equation}\label{htwomore}
h_{n,k}=n^k\,\prod_{i=0}^{k-1}\left(\mu+\frac{n^2\,(n-1)}{2}\,i\right)
\end{equation}

\end{corollary}

\begin{proof}
The proof follows from Proposition \ref{kernels} and (\ref{h}).
\end{proof}
 
\begin{corollary}\label{Gs} The functions $G_n$ appearing in (\ref{G0}) are given by

\begin{equation}\label{FIPone}
G_1(u,\mu)=e^{u\,\mu}
\end{equation}

\noindent  and for $n\geq2$

\begin{equation}\label{FIPtwomore}
G_n(u,\mu)=\left(1- \frac{n^3\,(n-1)}{2} \,u\right)^{ -\frac{2}{n^2\,(n-1)}\,\mu}=e^{-\frac{2}{n^2\,(n-1)}\,\mu \,\ln \left(1-\frac{n^3\,(n-1)}{2}\,u \right)}
\end{equation}

\bigskip

\noindent where  $\ln$ denotes logarithm with base $e$. 

\end{corollary}

\begin{proof} The proof follows from the fact that for $G_n$ given by (\ref{FIPone}) and (\ref{FIPtwomore}), in accordance with (\ref{G00}),  we have  

\[
\frac{\partial^k }{\partial u^k }\,G_n(u,\mu)|_{u=0}=n^k\,\prod_{i=0}^{k-1}\left(\mu+\frac{n^2\,(n-1)}{2}\,i\right)
\]

\end{proof}

\begin{corollary}\label{Ghats} The functions $\hat{G}_n$ appearing in (\ref{T}) are given by

\begin{equation}
\hat{G}_1(u)=u
\end{equation}

\noindent  and for $n\geq2$

\begin{equation}
\hat{G}_n(u)=-\frac{2}{n^2\,(n-1)}\,\ln \left(1-\frac{n^3\,(n-1)}{2}\,u \right)
\end{equation}

\end{corollary}

\begin{proof} The proof follows directly from Corollary \ref{Gs}.

\end{proof}

\begin{corollary}\label{FIP} The  $\mathcal{F}_n$ inner products  are given by

\begin{equation}\label{Gone}
\langle \psi_1(f),\psi_1(g)\rangle_1=e^{\int_{\mathbb{R}}\, \bar f(t)\,g(t)\,dt }
\end{equation}

\noindent  and for $n\geq2$

\begin{equation}\label{Gtwomore}
\langle \psi_n(f),\psi_n(g)\rangle_n=e^{ -\frac{2}{n^2\,(n-1)}\int_{\mathbb{R}}\, \ln\left(1-\frac{n^3\,(n-1)}{2}  \,\bar f(t)\,g(t)\right)\,dt}
\end{equation}

\bigskip

\noindent where $|f(t)|< \frac{1}{n}\,\sqrt{\frac{2}{n\,(n-1)}}$ and $|g(t)|<\frac{1}{n}\,\sqrt{\frac{2}{n\,(n-1)}}$.

\end{corollary}

\begin{proof} The proof follows from  (\ref{inner}) and Corollary \ref{Gs}.
\end{proof}

\noindent The function $G_1$ of (\ref{FIPone}) and the Fock space inner product (\ref{Gone}) are  associated with the Heisenberg-Weyl algebra and the quantum stochastic calculus of \cite{HudPa84}. For $n=2$ the function $G_n$ of ({\ref{FIPtwomore}) and the associated Fock space inner product (\ref{Gtwomore}) have appeared in the study of the Finite-Difference algebra and the Square of White Noise algebra in \cite{AS00}, \cite{ASk00}, \cite{Bou91}, and \cite{pfein}. The functions $G_n$  of ({\ref{FIPtwomore}) can also be found in Proposition 5.4.2 of  Chapter 5 of \cite{fein}.

\begin{definition} The $n$-th order TRHPWN Fock space $\mathcal{F}_n$ is the Hilbert space completion of the linear span of the exponential vectors  $\psi_n(f)$ of (\ref{expvect}) under the inner product $\langle\cdot, \cdot \rangle_n$ of Corollary \ref{FIP}.  The full TRHPWN Fock space $\mathcal{F}$ is the direct sum of the $\mathcal{F}_n$'s.
\end{definition}

\subsection{Fock representation of  the TRHPWN operators }

\begin{proposition}\label{genericops} For all test functions $f:=\sum_i\,a_i\,\chi_{I_i}$ and $g:=\sum_i\,b_i\,\chi_{I_i}$  with  $ I_i \cap I_j= \oslash $ for $i\neq j$, and for all $n\geq1$

\begin{equation}\label{B0n}
B^0_n(f)\,\psi_n(g)=n\,\int_{\mathbb{R}}\,f(t)\,g(t)\,dt\,\,\,\psi_n(g)+\frac{n^3\,(n-1)}{2}\,\frac{\partial}{\partial\,\epsilon}|_{\epsilon=0}\,\psi_n(g+\epsilon\,f\,g^2)
\end{equation}

\begin{equation}\label{Bn0}
B_0^n(f)\,\psi_n(g)=\frac{\partial}{\partial\,\epsilon}|_{\epsilon=0}\,\psi_n(g+\epsilon\,f)
\end{equation}

\end{proposition}

\begin{proof}  By (\ref{expvect}), the fact that  $\lbrack B^0_n(\chi_{I_i}), e^{B^n_0(\chi_{I_j})}\rbrack=0$ whenever $ I_i \cap I_j= \oslash $,  and by Lemma  \ref{ltnv}  we have

\begin{eqnarray*}
&&B_n^0(f)\, \psi_n(g) =\sum_{i=1}^m\,a_i\,B_n^0(\chi_{I_i})\,\prod_{j=1}^m\,e^{b_j\,B^n_0(\chi_{I_j})}\,\Phi \\
\noalign{\vskip .10 true in}
&& \quad =\sum_{i=1}^m\,a_i\,\prod_{j=1}^m\,B_n^0(\chi_{I_i})\,e^{b_j\,B^n_0(\chi_{I_j})}\,\Phi \\
&& \quad =\sum_{i=1}^m\,a_i\,\left( \prod_{  \stackrel{j=1}{j\neq i}  }^m\,e^{b_j\,B^n_0(\chi_{I_j})} \right)\,B_n^0(\chi_{I_i})\,e^{b_i\,B^n_0(\chi_{I_i})}\,\Phi \\
&& \quad= \sum_{i=1}^m\,a_i\,\left( \prod_{  \stackrel{j=1}{j\neq i}  }^m\,e^{b_j\,B^n_0(\chi_{I_j})} \right)\,\sum_{k=0}^{\infty}\,\frac{ b_i^k }{  k! }  \,B_n^0(\chi_{I_i})\,\left(B^n_0(\chi_{I_i})\right)^k\,\Phi 
\quad
\end{eqnarray*}

\begin{eqnarray*}
&& \quad =\sum_{i=1}^m\,a_i\,\left( \prod_{  \stackrel{j=1}{j\neq i}  }^m\,e^{b_j\,B^n_0(\chi_{I_j})} \right)\,\sum_{k=0}^{\infty}\,\frac{ b_i^k }{  k! }  \,  n\,k\,\left( \mu(I_i)+(k-1)\, \frac{n^2\,(n-1)}{2} \right)\,
\left( B^n_0(\chi_{I_i})\right)^{k-1} \,\Phi \\
&& \quad =\sum_{i=1}^m\,a_i\,\left( \prod_{  \stackrel{j=1}{j\neq i}  }^m\,e^{b_j\,B^n_0(\chi_{I_j})} \right)\,\sum_{k=1}^{\infty}\,\frac{ b_i^k }{  (k-1)! }  \, n\,\mu(I_i)\,
\left( B^n_0(\chi_{I_i})\right)^{k-1} \,\Phi \\
&& \quad +\sum_{i=1}^m\,a_i\,\left( \prod_{  \stackrel{j=1}{j\neq i}  }^m\,e^{b_j\,B^n_0(\chi_{I_j})} \right)\,\sum_{k=2}^{\infty}\,\frac{ b_i^k }{  (k-2)! }  \,\frac{n^3\,(n-1)}{2} \,\left( B^n_0(\chi_{I_i})\right)^{k-1} \,\Phi \\
&& \quad =n\,\sum_{i=1}^m\,a_i\,b_i\,\,\mu(I_i)\,\left( \prod_{  \stackrel{j=1}{j\neq i}  }^m\,e^{b_j\,B^n_0(\chi_{I_j})} \right)\, e^{b_i\,B^n_0(\chi_{I_i})}  \, \Phi \\
&& \quad +\frac{n^3\,(n-1)}{2} \,\sum_{i=1}^m\,a_i\,b_i^2\,B^n_0(\chi_{I_i})\,\left( \prod_{  \stackrel{j=1}{j\neq i}  }^m\,e^{b_j\,B^n_0(\chi_{I_j})} \right)\, e^{b_i\,B^n_0(\chi_{I_i})}   \,\Phi \\
&& \quad =n\,\sum_{i=1}^m\,a_i\,b_i\,\,\mu(I_i)\,\left( \prod_{ j=1}^m\,e^{b_j\,B^n_0(\chi_{I_j})} \right)\,\Phi 
\quad
\end{eqnarray*}

\begin{eqnarray*}
&& \quad +\frac{n^3\,(n-1)}{2} \,\sum_{i=1}^m\,a_i\,b_i^2\,B^n_0(\chi_{I_i})\,   e^{b_i\,B^n_0(\chi_{I_i})} \,\left( \prod_{  \stackrel{j=1}{j\neq i}    }^m\,e^{b_j\,B^n_0(\chi_{I_j})} \right)\,\Phi \\
&&  = n\,\int_{\mathbb{R}}\,f(t)\,g(t)\,dt\,\,\,\psi_n(g)  +\frac{n^3\,(n-1)}{2} \,\sum_{i=1}^m  \, \frac{\partial}{\partial\,\epsilon}|_{\epsilon=0}\,e^{( \epsilon \, a_i\,b_i^2 +b_i )\, B^n_0(\chi_{I_i})}\,\left( \prod_{  \stackrel{j=1}{j\neq i} }^m \,e^{b_j\,B^n_0(\chi_{I_j})} \right)  \,\Phi \\
\noalign{\vskip .10 true in}
&& \quad =n\,\int_{\mathbb{R}}\,f(t)\,g(t)\,dt\,\,\,\psi_n(g)  +\frac{n^3\,(n-1)}{2} \,\frac{\partial}{\partial\,\epsilon}|_{\epsilon=0}\,\left( \prod_{  i=1 }^m \,e^{ ( \epsilon \, a_i\,b_i^2 +b_i )   \,B^n_0(\chi_{I_i})} \right)  \,\Phi \\
&& \quad =  n\,\int_{\mathbb{R}}\,f(t)\,g(t)\,dt\,\,\,\psi_n(g)+\frac{n^3\,(n-1)}{2}\,\frac{\partial}{\partial\,\epsilon}|_{\epsilon=0}\,\psi_n(g+\epsilon\,f\,g^2)
\quad
\end{eqnarray*}

\noindent To prove (\ref{Bn0}) we notice that for $n=1$ (\ref{B0n}) yields

\[
B^0_1(f)\,\psi_1(g)=\int_{\mathbb{R}}\,f(t)\,g(t)\,dt\,\,\,\psi_1(g)
\]

\noindent i.e., $B^0_1(f)=A(f)$ where $A(f)$ is the annihilation operator of Hudson-Parthasarathy calculus (cf. \cite{HudPa84}) and so 

\[
B_0^1(f)\,\psi_1(g)=A^{\dagger}(f)\,\psi_1(g)=\frac{\partial}{\partial\,\epsilon}|_{\epsilon=0}\,\psi_1(g+\epsilon\,f)
\]

\noindent where $A^{\dagger}(f) $ is the creation operator of Hudson-Parthasarathy calculus thus proving (\ref{Bn0}) for $n=1$. To prove (\ref{Bn0}) for $n\geq 2$ we notice that by the duality condition (\ref{inv})  for all test functions $f, g, \phi$

\begin{eqnarray*}
&&\langle B^n_0(f)\,\psi_n(\phi), \psi_n(g)\rangle_n =   \langle \psi_n(\phi), B^0_n(\bar f)\,\psi_n(g)\rangle_n \\
\noalign{\vskip .12 true in}
&& \quad =  n\,\int_{\mathbb{R}}\,\bar f(t)\,g(t)\,dt\,\, \langle \psi_n(\phi), \psi_n(g)\rangle_n  + \frac{n^3\,(n-1)}{2}\,\frac{\partial}{\partial\,\epsilon}|_{\epsilon=0}\, \langle \psi_n(\phi), \psi_n(g+\epsilon\,\bar f\,g^2) \rangle_n\\
&& \quad = n\,\int_{\mathbb{R}}\,\bar f(t)\,g(t)\,dt\,\,  \langle \psi_n(\phi), \psi_n(g)\rangle_n    \\
 && \quad +\frac{n^3\,(n-1)}{2} \,\frac{\partial}{\partial\,\epsilon}|_{\epsilon=0}\,   e^{ -\frac{2}{n^2\,(n-1)}\int_{\mathbb{R}}\, \ln\left(1-\frac{n^3\,(n-1)}{2}  \,\bar \phi(t)\,(   g+ \epsilon \,\bar f \,g^2   )(t)\right)\,dt}  \\
&& \quad=   n\,\int_{\mathbb{R}}\,\bar f(t)\,g(t)\,dt\,\, \langle \psi_n(\phi), \psi_n(g)\rangle_n    \\
&& \quad +\frac{n^3\,(n-1)}{2} \,\, \langle \psi_n(\phi), \psi_n(g)\rangle_n   \,\left(  -\frac{2}{n^2\,(n-1)}\, \int_{\mathbb{R}}\,\frac{ -\frac{n^3\,(n-1)}{2}  \,\bar \phi \, \bar f \, g^2 }{ 1-\frac{n^3\,(n-1)}{2}  \,\bar \phi \, g } (t) \,dt \right)\\
&& \quad=  \left(   n\,\int_{\mathbb{R}}\,\bar f(t)\,g(t)\,dt+ \frac{n^4\,(n-1)}{2}\,  \int_{\mathbb{R}}\,\frac{ \bar \phi \, \bar f \, g^2 }{ 1-\frac{n^3\,(n-1)}{2}  \,\bar \phi \, g } (t) \,dt   \right)\,\langle \psi_n(\phi), \psi_n(g)\rangle_n  \\
 && \quad= n\,   \int_{\mathbb{R}}\,\frac{ \bar f \, g }{ 1-\frac{n^3\,(n-1)}{2}  \,\bar \phi \, g } (t) \,dt      \, \,\langle \psi_n(\phi), \psi_n(g)\rangle_n\\
 && \quad= \frac{\partial}{\partial\,\epsilon}|_{\epsilon=0}\,   e^{ -\frac{2}{n^2\,(n-1)}\int_{\mathbb{R}}\, \ln\left(1-\frac{n^3\,(n-1)}{2}  \,(   \bar \phi+ \epsilon \,\bar f  )(t)\,g(t)\right)\,dt}\\
 && \quad=\frac{\partial}{\partial\,\epsilon}|_{\epsilon=0}\, \langle \psi_n(\phi+\epsilon \,f), \psi_n(g)\rangle_n  \\
 && \quad= \langle \frac{\partial}{\partial\,\epsilon}|_{\epsilon=0}\,\psi_n(\phi+\epsilon \,f), \psi_n(g)\rangle_n  
\quad
\end{eqnarray*}

\noindent which implies (\ref{Bn0}).

\end{proof}

\begin{corollary}\label{number} For all $n\geq 1$ and  test functions 
$f,g,h$

\begin{eqnarray}\label{Bnk} 
&&B_{n-1}^{n-1}(f\,g)\,\psi_n(h)= \frac{1}{n}\, \int_{\mathbb{R}}\,  f (t)\, g(t) \,  \,\psi_n(h)\\
\noalign{\vskip .12 true in}
&&\quad+\frac{n\,(n-1)}{2}  \, \frac{\partial^2}{\partial\,\epsilon\,\,\partial\,\rho}|_{\epsilon=\rho=0}\,\left( \psi_n(h+\epsilon\,g+\rho\,\,f\,(h+\epsilon\,g)^2)-\psi_n(h+  \epsilon\,f\,h^2+\rho\,g)  \right)\nonumber
\quad
\end{eqnarray}

\end{corollary}

\begin{proof} 

\begin{eqnarray*}
&&B_{n-1}^{n-1}(f\,g)\,\psi_n(h)= \frac{1}{n^2}\,\lbrack B_{n}^0(f),B^{n}_0(g) \rbrack\,\psi_n(h)\\
\noalign{\vskip .12 true in}
&& \quad=\frac{1}{n^2}\,\left( B_{n}^0(f)\,B^{n}_0(g) -B^{n}_0(g)\,B_{n}^0(f)   \right)\,\psi_n(h)\\
&& \quad=\frac{1}{n^2}\,( B_{n}^0(f)\, \frac{\partial}{\partial\,\epsilon}|_{\epsilon=0}\,\psi_n(h+\epsilon\,g)-   B^{n}_0(g)\,(  n\,\int_{\mathbb{R}}\,f(t)\,h(t)\,dt\,\,\,\psi_n(h)\\
&&\quad+\frac{n^3\,(n-1)}{2}\,\frac{\partial}{\partial\,\epsilon}|_{\epsilon=0}\,\psi_n(h+\epsilon\,f\,h^2) ) )\\
&& \quad=\frac{1}{n^2}\,\frac{\partial}{\partial\,\epsilon}|_{\epsilon=0}\,B_{n}^0(f)\,\psi_n(h+\epsilon\,g)-  \frac{1}{n}\,\int_{\mathbb{R}}\,f(t)\,h(t)\,dt\,\,\,B^{n}_0(g)\,\psi_n(h)\\
&&\quad-\frac{n\,(n-1)}{2}\,\frac{\partial}{\partial\,\epsilon}|_{\epsilon=0}\,B^{n}_0(g)\,\psi_n(h+\epsilon\,f\,h^2)  \\
&&\quad =\frac{1}{n^2}\,\frac{\partial}{\partial\,\epsilon}|_{\epsilon=0}\,( n\,\int_{\mathbb{R}}\,f(t)\,(h+\epsilon \,g)(t)\,dt\,\,\,\psi_n(h+\epsilon \, g)\\
&&\quad +\frac{n^3\,(n-1)}{2}\,\frac{\partial}{\partial\,\rho}|_{\rho=0}\,\psi_n(h+\epsilon\,g+\rho\,\,f\,(h+\epsilon \, g)^2) )\\
&&\quad-\frac{1}{n}\,\int_{\mathbb{R}}\,f(t)\,h(t)\,dt\,\,\frac{\partial}{\partial\,\epsilon}|_{\epsilon=0}\,\psi_n(h+\epsilon \, g)-\frac{n\,(n-1)}{2}\,\frac{\partial}{\partial\,\epsilon}|_{\epsilon=0}\,\frac{\partial}{\partial\,\rho}|_{\rho=0}\,\psi_n(h+\epsilon\,f\,h^2+\rho\,g) \\
&&\quad=\frac{1}{n}\,\left(  \int_{\mathbb{R}}\,f(t)\,g(t)\,dt \,\,\,\psi_n(h) +\int_{\mathbb{R}}\,f(t)\,h(t)\,dt \,\,\,\frac{\partial}{\partial\,\epsilon}|_{\epsilon=0}\,  \psi_n(h+\epsilon \, g)  \right)\\
&&\quad +\frac{n\,(n-1)}{2}\, \frac{\partial^2}{\partial\,\epsilon\,\partial\,\rho}|_{\epsilon=\rho=0}   \,\psi_n(h+\epsilon\,g+\rho\,\,f\,(h+\epsilon \, g)^2) \\
&&\quad-\frac{1}{n}\,\int_{\mathbb{R}}\,f(t)\,h(t)\,dt\,\,\frac{\partial}{\partial\,\epsilon}|_{\epsilon=0}\,\psi_n(h+\epsilon \, g)\\
&&\quad-\frac{n\,(n-1)}{2}\,\frac{\partial^2}{\partial\,\epsilon\,\partial\,\rho}|_{\epsilon=\rho=0}\,\psi_n(h+\epsilon\,f\,h^2+\rho\,g) \\
&&\quad=\frac{1}{n}\, \int_{\mathbb{R}}\,f(t)\,g(t)\,dt \,\,\,\psi_n(h)\\
&&\quad+\frac{n\,(n-1)}{2}\,\frac{\partial^2}{\partial\,\epsilon\,\partial\,\rho}|_{\epsilon=\rho=0}\,\left( \psi_n(h+\epsilon\,g+\rho\,\,f\,(h+\epsilon \, g)^2)      -\psi_n(h+\epsilon\,f\,h^2+\rho\,g) \right)
\quad
\end{eqnarray*}

\end{proof}

\noindent Using the method described in Corollary \ref{number}, i.e., using the prescription

\[
 B^{n+N-1}_{k+K-1}( g f):=\frac{1}{ k\,N- K\,n }\, \left( B^n_k(g)\,B^N_K(f)-B^N_K(f)\,B^n_k(g)\right)
\]

\noindent and suitable linear combinations,  we obtain the representation of the $B^x_y$ (and therefore of the RHPWN and Virasoro--Zamolodchikov--$w_{\infty}$ commutation relations) on the appropriate Fock  space $\mathcal{F}_n$.

\section{Classical stochastic processes on $\mathcal{F}_n$}

\begin{definition}\label{cl} A quantum stochastic process $x=\{x(t)\,/\,t\geq 0\}$ is a family of Hilbert space operators. Such a process is said to be classical if for all $t, s \geq 0$,  $x(t)=x(t)^*$ and
  $[x(t),x(s)]:=x(t)\,x(s)-x(s)\,x(t)=0$. 
\end{definition}

\begin{proposition} Let $m>0$ and  let a quantum stochastic process $x=\{x(t)/\,t\geq0\}$ be defined by

\begin{eqnarray}
&x(t):=\sum_{n, k \in \Lambda } c_{n,k}\,B_k^n(t)&
\end{eqnarray}

\noindent where $c_{n,k}\in\mathbb{C}-\{0\}$, $\Lambda$ is a finite subset of $\{0,1,2,...\}$ and

\[
B_k^n(t):=B_k^n(\chi_{[0,t]})\in \mathcal{F}_m
\]

\noindent If  for each $n,k\in \Lambda $

\begin{eqnarray}
&c_{n,k}=\bar{c}_{k,n}&
\end{eqnarray}

\noindent then the process $x=\{x(t)\,/\,t\geq0\}$ is classical.

\end{proposition}

\begin{proof} 

 By (\ref{inv}), $x(t)=x^*(t)$ for all $t \geq 0$. Moreover, by (\ref{r2}),   $[x(t), x(s)]=0$ for all $t,s\geq0$ since each term of the form $c_{N,K}\,c_{n,k}\,[B^N_K(t), B^n_k(s)]$ is canceled out by the corresponding term of the form $c_{n,k}\,c_{N,K}\,[B^n_k(t), B^N_K(s)]$.  Thus the process $x=\{x(t)\,/\,t\geq0\}$ is classical.

\end{proof}

In the remaining of this section we will study the classical process $x=\{x(t)\,/\,t\geq 0\}$ whose Fock representation as a  family of operators on $\mathcal{F}_n$ is

\[
x(t):=B^n_0(t)+ B^0_n(t) 
\]

\noindent  By Proposition \ref{genericops} 

\begin{eqnarray}
B^n_0(t)\,\psi_n(g)&=&\frac{\partial}{\partial\,\epsilon}|_{\epsilon=0}\,\psi_n(g+\epsilon\,\chi_{[0,t]})\\
B_n^0(t)\,\psi_n(g)&=&n\,\int_0^t\,g(s)\,ds\,\,\,\psi_n(g)+\frac{n^3\,(n-1)}{2}\,\frac{\partial}{\partial\,\epsilon}|_{\epsilon=0}\,\psi_n(g+\epsilon \,\chi_{[0,t]}\,g^2)
\end{eqnarray}

\noindent In particular, for $g=0$

\begin{eqnarray}
B^n_0(t)\,\psi_n(0)&=&\frac{\partial}{\partial\,\epsilon}|_{\epsilon=0}\,\psi_n(\epsilon\,\chi_{[0,t]})\\
B_n^0(t)\,\psi_n(0)&=&0
\end{eqnarray}

\begin{lemma}[Splitting formula]\label{splitting} Let $s\in\mathbb{R}$. Then for $n=1$

\begin{equation}\label{HWsplform}
e^{s\,(B^1_0+B^0_1)}\,\Phi=e^{\frac{s^2}{2}\,\mu}e^{s\,B^1_0}\,\Phi
\end{equation}

\noindent and for $n\geq 2$

\begin{equation}\label{splform}
e^{s\,(B^n_0+B^0_n)}\,\Phi=\left(  \sec \left(\sqrt{\frac{n^3\,(n-1)}{2}}\,s    \right) \right)^{ \frac{2\,n\,\mu}{n^3\,(n-1)}  } \,e^{ \sqrt{\frac{2}{n^3\,(n-1)}}\, \tan\left(\sqrt{\frac{n^3\,(n-1)}{2}}\,s\right) \,B^n_0}\,\Phi
\end{equation}

\end{lemma} 

\begin{proof}  We will use  the ``differential method''  of  Proposition 4.1.1, Chapter 1  of \cite{fein}. So let

\begin{equation}\label{E}
E\,\Phi:=e^{s\,(B^n_0+B^0_n)}\,\Phi:=e^{V(s)\,B^n_0}\,e^{W(s)}\,\Phi
\end{equation}

\noindent where $W, V$ are real-valued functions with $W(0)=V(0)=0$. Then,

\begin{equation}\label{E1}
\frac{\partial}{\partial\,s}\,E\,\Phi=(B^n_0+B^0_n)\,E\,\Phi=B^n_0\,E\,\Phi+B^0_n\,E\,\Phi
\end{equation}

\noindent By Lemma \ref{ltnv} we have

\begin{eqnarray*}
&&B^0_{n}\,E\,\Phi=B^0_n\, e^{V(s)\,B^n_0}\,e^{W(s)}\,\Phi=e^{W(s)}\,B^0_n\, e^{V(s)\,B^n_0}\,\Phi\\
\noalign{\vskip .12 true in}
&&\quad=e^{W(s)}\,\sum_{k=0}^{\infty}\,\frac{V(s)^k}{k!}\,B^0_n\,(B^n_0)^k\,\Phi\\
&&\quad=e^{W(s)}\,\sum_{k=0}^{\infty}\,\frac{V(s)^k}{k!}\,n\,k\,\left(\mu+(k-1)\,\frac{n^2\,(n-1)}{2}\right)\,(B^n_0)^{k-1}\,\Phi\\
&&\quad (n\,\mu\,V(s)+\frac{n^3\,(n-1)}{2}\,V(s)^2\,B^n_0)\,e^{V(s)\,B^n_0}\,e^{W(s)}\,\Phi\\
&&\quad (n\,\mu\,V(s)+\frac{n^3\,(n-1)}{2}\,V(s)^2\,B^n_0)\,E\,\Phi
\quad
\end{eqnarray*}

\noindent Thus (\ref{E1}) becomes

\begin{equation}\label{E2}
\frac{\partial}{\partial\,s}\,E\,\Phi=\left(B^n_0+n\,\mu\,V(s)+\frac{n^3\,(n-1)}{2}\,V(s)^2\,B^n_0   \right)\,E\,\Phi
\end{equation}

\noindent From (\ref{E}) we also have

\begin{equation}\label{E3}
\frac{\partial}{\partial\,s}\,E\,\Phi=\left(V^{\prime}(s)\,B^n_0+  W^{\prime}(s) \right)\,E\,\Phi
\end{equation}

\noindent From (\ref{E2}) and  (\ref{E3}),  by equating coefficients of $1$ and $B^n_0$,  we  have
 
\begin{eqnarray}
W^{\prime}(s)&=&n\,\mu\,V(s)\\
V^{\prime}(s)&=&1+\frac{n^3\,(n-1)}{2}\,V(s)^2 \mbox{  (Riccati equation)}\label{Riccati}
\end{eqnarray}

\noindent For $n=1$ we find $V(s)=s$ and $W(s)=\frac{s^2}{2}\,\mu$. For $n\geq2$ by separating the variables we find

\[
V(s)=\sqrt{\frac{2}{n^3\,(n-1)}}\, \tan\left(\sqrt{\frac{n^3\,(n-1)}{2}}\,s\right)
\]

\noindent and so 

\[
W(s)=-\frac{2\,n\,\mu}{n^3\,(n-1)}\,\ln\left( \cos \left( \sqrt{\frac{n^3\,(n-1)}{2}}\,s   \right)  \right)
\]

which implies that

\[
e^{W(s)}=\left(  \sec \left(\sqrt{\frac{n^3\,(n-1)}{2}}\,s    \right) \right)^{ \frac{2\,n\,\mu}{n^3\,(n-1)}  } 
\]

\noindent thus completing the proof.

\end{proof}

In the theory of Bernoulli systems and the Fock representation of finite-dimensional Lie algebras (cf. Chapter 5 of \cite{fein})   the Riccati equation (\ref{Riccati}) has the general  form

\[
V^{\prime}(s)=1+2\,\alpha\,V(s)+\beta\,V(s)^2
\]

\noindent and the values of $\alpha$ and $\beta$ determine the underlying classical probability distribution and the associated special functions. For example, for $\alpha=1-2\,p$ and $\beta=-4\,p\,q$ we have the binomial process and the Krawtchouk polynomials, for $\alpha=p^{-1}-\frac{1}{2}$ and $\beta=q\,p^{-2}$ we have the negative binomial process and the Meixner polynomials,  for $\alpha \neq 0$ and $\beta=0$ we have the Poisson process and the Poisson-Charlier polynomials,  for $\alpha^2=\beta$ we have the exponential process and the Laguerre  polynomials, for $\alpha=\beta=0$ we have Brownian motion with moment generating function $e^{\frac{s^2}{2}\,t}$ and associated special functions the Hermite polynomials, and for $\alpha^2-\beta <0$ we have the continuous binomial and Beta processes (cf.  Chapter 5 of \cite{fein} and also \cite{Feller} )
 with moment generating function $(\sec s)^t$  and associated special functions the Meixner-Pollaczek polynomials.
In the infinite-dimensional TRHPWN case the underlying classical probability distributions are given in the following.

\begin{proposition}[Moment generating functions] \label{mgf}For all $s\geq0$

\begin{equation}\label{eqmgf1}
\langle  e^{s\,(B^1_0(t)+B^0_1(t))}\,\Phi, \Phi \rangle_1=e^{\frac{s^2}{2}\,t}
\end{equation}

\noindent i.e., $\{B^1_0(t)+B^0_1(t)\,/\,t\geq0\}$ is Brownian motion (cf. \cite{fein}, \cite{HudPa84} ) while for $n\geq2$

\begin{equation}\label{eqmgf}
\langle  e^{s\,(B^n_0(t)+B^0_n(t))}\,\Phi, \Phi \rangle_n=\left(  \sec \left(\sqrt{\frac{n^3\,(n-1)}{2}}\,s    \right) \right)^{ \frac{2\,n\,t}{n^3\,(n-1)}  } 
\end{equation}

\noindent i.e., $\{B^n_0(t)+B^0_n(t)\,/\,t\geq0\}$ is for each $n$ a  continuous binomial/Beta process (see Appendix) 
\end{proposition}

\begin{proof}  The proof follows from Lemma \ref{splitting},  $\mu([0,t])=t$,  and the fact that  for all  $n\geq1$ we have $B^0_n(t)\,\Phi=0$.

\end{proof}

\section{Appendix: The continuous Binomial and Beta  Processes}

Let 

\[
b(n,k)=\binom{n}{k}\,x^k\,(1-x)^{n-k}\,\,\,;\,\,\,n,k\in\{0,1,2,...\},\,\, n\geq k,\,\, x\in (0,1)
\]

\noindent be the standard Binomial distribution. Using the Gamma function we can analytically extend from $n,k\in\{0,1,2,...\}$ to $z,w\in\mathbb{C}$ with $\Re\,z\,\geq\,\Re\,w\,>\,-1$  and we have

\begin{eqnarray*}
b(z,w)&=&\frac{\Gamma(z+1)}{\Gamma(z-w+1)\,\Gamma(w+1)}\,x^w\,(1-x)^{z-w}\\
&=&\frac{1}{z+1}\,\frac{\Gamma(z+2)}{\Gamma(z-w+1)\,\Gamma(w+1)}\,x^w\,(1-x)^{z-w} \\
&=&\frac{1}{z+1}\,\frac{\Gamma(z+2)}{\Gamma(z-w+1)\,\Gamma(w+1)}\,x^{(w+1)-1}\,(1-x)^{(n-w+1)-1}\\
&=&\frac{1}{z+1}\,\beta(w+1,\,z-w+1)
\end{eqnarray*}

\noindent where $\beta(w+1,\,z-w+1)$ is the analytic continuation to $\Re\,a \,>\,0$ and $\Re\,c\, >\,0$  of the 
standard Beta distribution

\[
\beta(a,c)=\frac{\Gamma(a+c)}{\Gamma(a)\,\Gamma(c)}\,x^{a-1}\,(1-x)^{c-1}\,\,\,;\,\,\,a>0\,,\,c>0
\]

\begin{proposition}\label{sec}  For each $t>0$ let $X_t$ be a random variable with distribution given by the density

\[
p_t(x)=\frac{2^{t-1}}{2\,\pi}\,\beta(\frac{t+i\,x}{2},\frac{t-i\,x}{2})
\]

\noindent Then the moment generating funcion of $X_t$ is 

\begin{equation}\label{m}
\langle e^{s\,X_t} \rangle :=\int_{-\infty}^{\infty}\, e^{s\,x} \,p_t(x)\,dx=(\sec s)^t\,\,\,\,;\,\,\,\,\forall \, t>0,\,s\in\mathbb{R}
\end{equation}

\end{proposition}

\begin{proof}
 See  Proposition 4.1.1, Chapter 5 of \cite{fein}.
\end{proof}

\begin{corollary} With $X_t$ and $p_t$ as in Proposition \ref{sec}, let

\[
Y_t:= \sqrt{\frac{n^3\,(n-1)}{2}}\,X_t
\]

\noindent Then the moment generating funcion of $Y_t$  with respect to  the density

\[
q_t:=p_{\frac{2\,n}{n^3\,(n-1)}\,t}
\]

\noindent where $n\in\{1,2,...\}$, is 

\[
\langle e^{s\,Y_t} \rangle =\left(  \sec \left(\sqrt{\frac{n^3\,(n-1)}{2}}\,s    \right) \right)^{ \frac{2\,n\,t}{n^3\,(n-1)}  } 
\]

\end{corollary}

\begin{proof} Since $p_t$ is for each $t>0$ a probability density function 
we have

\[
\int_{-\infty}^{\infty}\,p_t(x)\,dx=1\,\,\,\,;\,\,\,\,\forall \, t>0
\]

\noindent and so for $t:=\frac{2\,n}{n^3\,(n-1)}\,t$

\[
\int_{-\infty}^{\infty}\,p_{ \frac{2\,n}{n^3\,(n-1)}\,t}(x)\,dx=1\,\,\,\,;\,\,\,\,\forall \, t>0
\]

\noindent i.e.,

\[
\int_{-\infty}^{\infty}\,q_t(x)\,dx=1\,\,\,\,;\,\,\,\,\forall \, t>0
\]

\noindent  so $q_t$ is for each $t>0$ a probability density function. Moreover,  letting $t:=\frac{2\,n}{n^3\,(n-1)}\,t$   and    $s:=\sqrt{\frac{n^3\,(n-1)}{2}}\,s$    in (\ref{m}) we obtain

\[
\int_{-\infty}^{\infty}\, e^{ s\,\sqrt{\frac{n^3\,(n-1)}{2}} \,x} \,q_t(x)\,dx=\left(\sec \left( \sqrt{\frac{n^3\,(n-1)}{2}}\,s  \right)\right)^{\frac{2\,n\,t}{n^3\,(n-1)}}
\]

\noindent which is precisely the moment generating function $\langle e^{s\,Y_t} \rangle $ of $Y_t$ with respect to $q_t$.

\end{proof}


\begin{thebibliography}{99}

\bibitem{ABIDAQP06} Accardi, L., Boukas, A.:  Renormalized higher powers of white noise (RHPWN) and conformal field theory,  {\em Infinite Dimensional Anal. Quantum  Probab. Related Topics} {\bf 9}, No. 3,  (2006) 353-360. 

\bibitem{ABIJMCS06} \bysame : The emergence of the Virasoro and $w_{\infty}$ Lie algebras through the renormalized higher powers of quantum white noise , {\em International Journal of Mathematics and Computer Science}, {\bf 1},  No.3, (2006) 315--342.

\bibitem{id} \bysame : Renormalized Higher Powers of White Noise and the Virasoro--Zamolodchikov--$w_{\infty}$ Algebra, submitted (2006), http://arxiv.org/hep-th/0610302.

\bibitem{ABCOSA06} \bysame : Lie algebras associated with the renormalized higher powers of white noise,  to appear in {\em Communications on Stochastic Analysis} (2006).

\bibitem{ABF06} Accardi, L., Boukas, A., Franz, U.: Renormalized powers of quantum white noise,  {\em Infinite Dimensional Analysis, Quantum Probability, and Related Topics}, {\bf 9}, No. 1, 129--147, (2006).

\bibitem{AFS02} Accardi, L.,  Franz, U., and  Skeide, M.: Renormalized 
squares of white noise and non-- Gaussian noises as Levy processes on real 
Lie algebras;  {\em Comm. Math. Phys.} {\bf 228}, No. 1, (2002) 
123--150.

\bibitem{ALV99} Accardi, L.,  Lu, Y.G.,  Volovich, I. V.:  White noise approach to classical and quantum stochastic calculi, {\em Lecture Notes of the Volterra International School of the same title}, Trento, Italy, 1999, Volterra Center preprint 375.

\bibitem{AS00}	Accardi, L., Skeide, M.: Hilbert module realization of the square of white noise and finite difference algebras, {\em Math. Notes} 68(5-6), 683--694, (2000).

\bibitem{ASk00} Accardi, L., Skeide, M.: On the relation of the square of white noise and the finite difference algebra,  {\em Infinite dimensional analysis, quantum probability and related topics}, vol. 3, no. 1, 185--189 (2000).

\bibitem{BK91} Bakas, I., Kiritsis, E.B.: Structure and representations of the $W_{\infty}$ algebra, {\em Prog. Theor. Phys. Supp.} {\bf 102} (1991) 15.

\bibitem{Bou91}Boukas, A.: An Example of a Quantum Exponential Process, {\em Mh. Math.}, 112, 209-215(1991).

\bibitem{pfein} Feinsilver, P. J.: Discrete analogues of the Heisenberg-Weyl algebra, \emph{Mh. Math.}, 104:89--108 (1987).

\bibitem{fein}  Feinsilver, P. J.,  Schott, R.: \emph{Algebraic structures and operator calculus. Volumes I and III}, Wiley, 1971.

\bibitem{Feller} Feller, W.: \emph{Introduction to probability theory and its applications. Volumes I and II}, Kluwer,  1993.

\bibitem{HudPa84}
Hudson R.L., Parthasarathy K.R.: Quantum Ito's formula and
stochastic evolutions, \emph{Comm. Math. Phys.} 93 (1984), 301--323.

\bibitem{K95}  Ketov, S. V.: {\em Conformal field theory}, World Scientific, 1995.

\bibitem{Z85} Zamolodchikov, A.B.:  Infinite additional symmetries in two-dimensional conformal quantum field theory, {\em Teo. Mat. Fiz.} {\bf 65} (1985), 347--359. 

\end{thebibliography}
\end{document}